\documentclass[12pt]{article}
\usepackage[english]{babel}
\usepackage{graphicx,epsfig}
\makeindex \textwidth 170mm \textheight 220mm \topmargin -5mm
\newcommand{\be}{\begin{equation}}
\newcommand{\ee}{\end{equation}}
\newcommand{\bea}{\begin{eqnarray}}
\newcommand{\eea}{\end{eqnarray}}

\def\rfr#1{eq. (\ref{#1})}

\def\eqi{\begin{equation}}
\def\eqf{\end{equation}}
\def\eqia{\begin{eqnarray}}
\def\eqfa{\end{eqnarray}}
\def\btab{\begin{tabular}}
\def\etab{\end{tabular}}
\def\bar{\begin{array}}
\def\ear{\end{array}}

\def\GR{General Relativity}
\def\grl{general relativistic}
\def\wfs{weak--field and slow--motion approximation}
\def\leti{Lense--Thirring}
\def\grc{gravitomagnetic}

\def\se{systematic error}

\def\zh{even zonal harmonics}

\def\gp{geopotential}

\def\lg{{\rm LAGEOS}}
\def\lgg{{\rm LAGEOS} II}

\def\lb#1{\label{#1}}
\def\pc{precession}
\def\nd{node}
\def\pg{perigee}
\def\nl{nodal}

\def\sa{semimajor axis}
\def\ec{eccentricity}
\def\ic{inclination}

\def\et{Earth}
\def\ef{effect}
\def\dt#1{\dot{#1}}

\def\st{satellite}
\def\lt{_{\rm{LT}}}

\begin{document}
\begin{titlepage}
\begin{flushright}
\today\\
BARI-TH/00\\
\end{flushright}
\vspace{.5cm}
\begin{center}
{\LARGE A new proposal for measuring the Lense--Thirring effect
with a pair of supplementary satellites in the gravitational field
of the Earth } \vspace{1.0cm}
\quad\\
{Lorenzo Iorio$^{\dag}$\\ \vspace{0.1cm}
\quad\\
{\dag}Dipartimento di Fisica dell' Universit{\`{a}} di Bari, via
Amendola 173, 70126, Bari, Italy\\ \vspace{0.2cm} } \vspace{0.2cm}

{\bf Abstract}\\
\end{center}

{\noindent \small In this letter we propose a new observable for
measuring the general relativistic Lense--Thirring effect with
artificial satellites in the gravitational field of the Earth. It
consists of the difference of the perigee rates of two satellites
placed in identical orbits with supplementary inclinations. As in
the well known LAGEOS--LARES project in which, instead, the sum of
the residuals of the nodal rates would be used, the proposed
observable would be able to cancel out the aliasing effect of the
classical even zonal perigee precessions induced by the oblateness
of the Earth. The possibility of using the already existing LAGEOS
II and a twin of its, to be launched, in a supplementary orbit is
briefly examined. While with the originally proposed LAGEOS--LARES
mission only the sum of the nodal rates could be used because the
perigee of LAGEOS is not a good observable, the implementation of
the proposed mission would allow to adopt both the sum of the
nodes and the difference of the perigees.}

{\noindent \small }
\end{titlepage} \newpage \pagestyle{myheadings} \setcounter{page}{1}
\vspace{0.2cm} \baselineskip 14pt

\setcounter{footnote}{0}
\setlength{\baselineskip}{1.5\baselineskip}
\renewcommand{\theequation}{\mbox{$\arabic{equation}$}}
\noindent

\section{Introduction}
In its \wfs\ \GR\ predicts that, among other things,  the orbit of
a test particle freely falling in the gravitational field of a
central spherical rotating body is affected by the so called \grc\
dragging of the inertial frames or
\leti\ \ef. More precisely,  the longitude of the ascending \nd\ $\Omega$ and
the argument of the \pg\ $\omega$ of the orbit \cite{ste} undergo
tiny \pc s \cite{ciuwhe, iorncb} (The original papers by Lense and
Thirring can be found in english translation in \cite{let} where the longitude of the perigee
$\varpi=\Omega+\omega$ is used instead of $\omega$ which is an angle counted in the osculating orbital plane
from the line of the nodes to the direction of the perigee.) \eqia
\dot\Omega \lt & = &
\frac{2GJ}{c^{2}a^{3}(1-e^{2})^{\frac{3}{2}}},\\
\dot\omega \lt & = &
-\frac{6GJ\cos{i}}{c^{2}a^{3}(1-e^{2})^{\frac{3}{2}}},\lb{perigeo}\eqfa
in which $G$ is the Newtonian gravitational constant, $J$ is the
proper angular momentum of the central body supposed spherically
symmetric and rigidly rotating, $c$ is the speed of light $in\
vacuum$, $a,\ e$ and $i$ are the \sa, the \ec\ and the \ic,
respectively, of the orbit of the test particle.

The first measurement of this \ef\ in the gravitational field of
the \et\ has been obtained by analyzing a suitable combination of
the laser-ranged data to the existing passive geodetic \st s \lg\
and \lgg\ \cite{ciuetal}. The observable \cite{ciu2} is a linear
trend with a slope of 60.2 milliarcseconds per year (mas/y in the
following) and includes the residuals of the nodes of \lg\ and
\lgg\ and the \pg\ of \lgg\footnote{The \pg\ of \lg\ was not used
because it introduces large observational errors due to the
smallness of the \lg\ \ec. \cite{ciu2} which amounts to 0.0045.}.
The
\leti\ precessions for the \lg\ satellites amount to \eqia
\dot\Omega\lt^{\rm LAGEOS}& = & 31\ \textrm{mas/y},\\
\dot\Omega\lt^{\rm LAGEOS\ II} & = & 31.5\ \textrm{mas/y},\\
\dot\omega\lt^{\rm LAGEOS} & = & 31.6\ \textrm{mas/y},\\
\dot\omega\lt^{\rm LAGEOS\ II} & = & -57\ \textrm{mas/y}. \eqfa
The claimed total relative accuracy of the measurement is $2\times 10^{-1}$ \cite{ciuetal}.

In this kind of experiment the major source of \se s is
represented by the aliasing trends due to the classical secular
precessions \cite{kau} of the \nd\ and the \pg\ induced by the
mismodelled \zh\ of the \gp\ $\delta J_2,\ \delta J_4,\ \delta
J_6,...$ Indeed, according to the present knowledge of the \et's
gravity field based on EGM96 model \cite{lem}, they amount to a
large part of the \grc\ precessions of interest, especially for
the first two even zonal harmonics. In the performed \lg\
experiment the adopted observable allowed for the cancellation of
the static and dynamical effects of $\delta J_2$ and $\delta J_4$.
The remaining higher degree even zonal harmonics affected the
measurement at a $12.9\%$ level, according to the covariance matrix of the 
EGM96 up to degree $l=20$.

In order to achieve a few percent accuracy, in \cite{ciu1} it was
proposed to launch a passive geodetic laser-ranged \st- the former
{\rm LAGEOS} III - with the same orbital parameters of \lg\ apart
from its inclination which should be supplementary to that of \lg.

This orbital configuration would be able to cancel out exactly the
classical \nl\ \pc s, which are proportional to $\cos i$, provided
that the observable to be adopted is the sum of the residuals of
the \nl\ \pc s of {\rm LAGEOS} III and LAGEOS \eqi
\delta\dt\Omega^{{\rm III}}+\delta\dt\Omega^{{\rm
I}}.\lb{lares}\eqf The relativistic signature would be a linear trend with a slope
of 62 mas/y. Later on the concept of the mission
slightly changed. The area-to-mass ratio of {\rm LAGEOS} III was
reduced in order to make less relevant the impact of the
non-gravitational perturbations  and the eccentricity  was
enhanced in order to be able to perform other \grl\ tests: the
LARES was born \cite{ciu3}.

Currently, the observable of the LAGEOS--LARES mission is under
revision in order to improve the obtainable accuracy
\cite{ioretal}.

The orbital parameters of \lg, \lgg\ and LARES are in Tab. 1.

\begin{table}[ht!]
\caption{Orbital parameters of \lg, \lgg\ and LARES.} \label{para}
\begin{center}
\begin{tabular}{lllll}
\noalign{\hrule height 1.5pt} Orbital parameter & \lg & \lgg &
LARES\\ \hline
$a$ (km) & 12,270 & 12,163 & 12,270\\
$e$ & 0.0045 & 0.014 & 0.04\\
$i$ (deg) & 110 & 52.65 & 70\\
\noalign{\hrule height 1.5pt}
\end{tabular}
\end{center}
\end{table}
\section{A new perigee--only observable}
The concept of a couple of satellites placed in identical orbits
with supplementary inclinations could be fruitfully exploited in
the following new way.

An inspection of \rfr{perigeo} and of the explicit expressions of
the rates of the classical perigee precessions induced by the even
zonal harmonics of the geopotential \cite{ior2} suggests to adopt
as observable the difference of the residuals of the perigee
precessions of the two satellites
\eqi\delta\dot\omega^{i}-\delta\dot\omega^{180^{\circ}-i},\lb{new}\eqf 
so to obtain a secular trend with a
certain slope in mas/y. Indeed, on one hand, the
Lense--Thirring perigee precessions depend on $\cos i$, contrary
to the nodal rates which are independent of the inclination, so
that, by considering the relativistic effect as an unmodelled
force entirely adsorbed in the residuals, in \rfr{new} they sum
up. On the other, it turns out that the classical even zonal
perigee precessions depend on even powers of $\sin i$ and on
$\cos^2 i$, so that they cancel out exactly in \rfr{new}. It may
be interesting to notice that the proposed observable of \rfr{new}
is insensitive to the other general relativistic feature which
affect the pericenter of a test body, i.e. the gravitoelectric
Einstein precession. Indeed, as it is well known \cite{ciuwhe}, it
does not depend on the inclination of the orbital plane.

In regard to a practical application of such idea, we note that
the LAGEOS--LARES mission would be unsuitable because the perigee
of LAGEOS is not a good observable due to the notable smallness of
the eccentricity of its orbit. For the sake of concreteness, we
could think about a LARES II which should be the supplementary
companion of LAGEOS II. In this case we would have a
gravitomagnetic trend with a slope of -115.2 mas/y (which is
almost twice that of the LAGEOS--LARES node--only mission).
Moreover, since the magnitude of the eccentricity of LAGEOS II is
satisfactory in order to perform relativistic measurements with
its perigee, the LARES II, contrary to the LAGEOS--LARES mission,
could be inserted in an orbit with the same eccentricity of that
of LAGEOS II. So, the cancellation of the classical secular
precessions would occur at a higher level than in the
LAGEOS--LARES node--only observable \cite{ioretal}. Of course, a
careful analysis of the time--dependent gravitational and,
especially, non--gravitational perturbations (see \cite{luc1} for
the radiative perturbations and \cite{luc2} for the thermal,
spin--dependent perturbations), to which the perigee is
particularly sensitive, contrary to the node, would be needed in
order to make clear if also for such perturbations some useful
cancellations may occur. Among the gravitational tidal
perturbations, fortunately, the most relevant components of the
semi--secular 18.6--year and the 9.3--year tides are even ($l=2$)
zonal ($m=0$) \cite{ior1}, so that they would be canceled out. The
attention should be focused on some insidious non--gravitational
thermal perturbations. Indeed, some of them can generate linear
perturbations, like the terrestrial Yarkovski--Rubincam effect,
which could mimic the relativistic signal. From very preliminary
investigations based on \cite{luc2} it seems that such terms would
cancel out because proportional to $\cos^2 i$, apart from other
characteristics of the satellites like the thermal lag angle
$\vartheta$ and the square of the component $S_z^2$ of the spin
vector along the $z$ axis of an inertial frame which could be
identically prepared for the two satellites. On the contrary,
according to \cite{luc1} many components of the radiative
perturbations would not cancel out because they depend on $\cos
i$. However, in regard to such other harmonic perturbations, even
if not cancelled out, their impact would be less dangerous
because, over long enough time spans, they could be viewed as
empirically fitted quantities and removed from the signal. An
optimal choice would be the launch of a couple of entirely new,
geodetic satellites in highly eccentric orbits with supplementary
inclinations to be carefully selected so to reduce the periods of
the time--varying perturbations which could affect the observable.
By the way, it should also be noticed that the implementation of
such a mission would provide us with, at least, two different and
complementary gravitomagnetic observables: $\delta\dot\Omega^i
+\delta\dot\Omega^{180^{\circ}-i}$ and $\delta\dot\omega^i
-\delta\dot\omega^{180^{\circ}-i}$. Instead, the originally
proposed LAGEOS--LARES mission would allow to use
$\delta\dot\Omega^i +\delta\dot\Omega^{180^{\circ}-i}$ only.
Moreover, the presence in orbit of LARES with LARES II or,
eventually, two entirely new satellites would strongly enhance the
possibility of fruitfully following the strategy of the
multisatellite combined residuals, as sketched in \cite{ioretal}.
More extensive and quantitative investigations can be found in \cite{iorperi,iorluc}
Of course, also a pair of drag-free satellites, although more expensive,
could be employed so to reduce dramatically the impact
of the non--gravitational perturbations.
\section{Conclusions}
In this letter we have proposed a new observable for the
measurement of the general relativistic Lense--Thirring effect on
the orbital motion of artificial satellites in the gravitational
field of the Earth. It consists of the difference of the perigee
rates of two satellites placed in identical orbits, except for the
inclinations which should be supplementary. Such proposal could be
implemented by means of a LARES II satellite with the same orbital
parameters of the already existing LAGEOS II, except for the
inclination, or by launching an entirely new couple of geodetic
satellites in highly eccentric orbits with carefully selected
supplementary inclinations. The latter choice could reduce the
periods of the non--gravitational perturbations, which have, in
general, a non negligible impact on the perigees of geodetic
satellites, affecting the proposed observable. Forthcoming
investigations of the time--dependent gravitational and
non--gravitational perturbations will make clear their role in the
sketched LAGEOS II--LARES II perigee--only mission. Of course, in
addition to the proposed perigee--only measurement, also the sum
of the nodes could be investigated, so to enforce and enlarge the
experimental basis for a detection of the elusive Lense--Thirring
effect. 
\section*{Acknowledgements}
I'm grateful to L. Guerriero for his support while at Bari and to
D.M. Lucchesi for his helpful and important informations on the
non--gravitational perturbations on LAGEOS II.

\end{document}